\documentclass[journal]{IEEEtran}  

\usepackage[utf8]{inputenc}
\usepackage{amsmath}
\usepackage{amssymb}
\usepackage{amsfonts}
\usepackage{tabularx}
\usepackage{array}
\usepackage{algpseudocode}
\usepackage{subcaption}
\usepackage{graphicx}
\usepackage{algorithm}
\usepackage{xcolor, soul}

\renewcommand{\hl}[1]{#1}
\newcommand{\bc}[1]{\textcolor{blue}{#1}}
\renewcommand{\bc}[1]{#1}

\usepackage{tikz}
\usepackage{textcomp}
\usepackage{hyperref}
\usepackage{lipsum}

\newcommand\copyrighttext{%
  \footnotesize \textcopyright 2024 IEEE. Personal use is permitted, but republication/redistribution requires IEEE permission.
This article has been accepted for publication in IEEE Transactions on Services Computing. This is the author's version which has not been fully edited and content may change prior to final publication. Citation information: DOI: \href{https://doi.org/10.1109/TSC.2024.3395919}{10.1109/TSC.2024.3395919}.
}
\newcommand\copyrightnotice{%
\begin{tikzpicture}[remember picture,overlay]
\node[anchor=south,yshift=755pt] at (current page.south) {\fbox{\parbox{\dimexpr\textwidth-\fboxsep-\fboxrule\relax}{\copyrighttext}}};
\end{tikzpicture}%
}

\title{Resource-aware Cyber Deception \bc{for \\ Microservice-based Applications}}
\author{Marco Zambianco, Claudio Facchinetti, Roberto Doriguzzi-Corin, Domenico Siracusa\\
\small{Center for Cybersecurity, Fondazione Bruno Kessler, Italy}
}
\begin{document}

\maketitle

\copyrightnotice

\begin{abstract}
 Cyber deception can be a valuable addition to traditional cyber defense mechanisms, especially for modern cloud-native environments with a fading security perimeter. However, pre-built decoys used in classical computer networks are not effective in detecting and mitigating malicious actors due to their inability to blend with the variety of applications in such environments. On the other hand, decoys cloning the deployed microservices of an application can offer a high-fidelity deception mechanism to intercept ongoing attacks within production environments. However, to fully benefit from this approach, it is essential to use a limited amount of decoy resources and devise a suitable cloning strategy to minimize the impact on legitimate services performance. 
 Following this observation, we formulate a non-linear integer optimization problem that maximizes the number of attack paths intercepted by the allocated decoys within a fixed resource budget. Attack paths represent the attacker's movements within the infrastructure as a sequence of violated microservices. We also design a heuristic decoy placement algorithm to approximate the optimal solution and overcome the computational complexity of the proposed formulation.
 We evaluate the performance of the optimal and heuristic solutions against other schemes that use local vulnerability metrics to select which microservices to clone as decoys. Our results show that the proposed allocation strategy achieves a higher number of intercepted attack paths compared to these schemes while requiring approximately the same number of decoys.
\end{abstract}

\section{Introduction}

Cloud-native applications running on highly-distributed environments with a variety of interfaces and microservices inevitably increase the attack surface of the infrastructure \cite{nadareishvili2016microservice}. Traditional cyber defense mechanisms, like intrusion detection systems (IDSs), rely on up-to-date knowledge of attacker tactics, techniques, and procedures (TTPs), which frequently evolve due to the multitude of technologies employed \cite{hannousse2021securing}. As a result, production applications are more exposed to zero-day attacks where deployed microservices can be exploited by malicious actors as a foothold to further spread within the target environment.

To strengthen security, cyber deception can be used as a proactive defense strategy by allocating decoys that resemble legitimate system components within the defender infrastructure to lure malicious actors into interacting with them \cite{jajodia2016cyber}. Decoys must appear realistic to engage the attacker and increase the likelihood of interaction. Traditionally, decoys are general-purpose (possibly vulnerable) applications that are created a-priori and shipped to customers and, as such, they might have little or no fit within the defender ecosystem \cite{zhang2021three}. Moreover, when prompted by an attacker they cannot ensure the same degree of similarity or interactivity level as the original applications \cite{duan2018conceal}\cite{ferguson2021examining}\cite{islam2021chimera}.
On the other hand, containerization techniques and cloud-native technologies provide the instruments to create high-fidelity and high-interaction decoys by cloning microservices of deployed applications with limited complexity \cite{nehme2019securing}. Unlike pre-built decoys, which require a non-negligible management overhead and are usually installed as standalone elements within the defender environment \cite{han2018deception}, deceptive microservices can be seamlessly deployed and monitored in production using cloud orchestration technologies like Kubernetes.

By blending this deception mechanism with the running microservices, defenders can detect and mitigate internal threats such as attacker lateral movements between microservices while gathering valuable information about the attacker TTPs. Moreover, using clones of organization's microservices as decoys reduces the chance that attackers will unveil the deception mechanism since they are more likely to believe that the decoys are legitimate system components. This belief is further enforced by the fact that replication of microservices is a well-known horizontal scaling technique adopted to accommodate variable workload conditions \cite{hasselbring2017microservice}. Indeed, the attacker might expect to find multiple copies of the same microservice within the system configuration.

To fully benefit from this deception approach, two main challenges must be addressed. Firstly, enterprises cannot afford a massive deployment of decoys within their infrastructure, due to limited computational resources, potential scarcity (e.g., in edge scenarios), and associated costs (capital and/or operational expenditure). A key requirement for a practical deception solution is therefore balancing the number of decoys based on incurred resource usage. Secondly, unlike traditional decoy allocation schemes that consider decoys as isolated from the real system in order to avoid any negative impact on the production assets, a deceptive ecosystem based on copies of existing microservices is intertwined with the real application deployment. Consequently, the decoy allocation strategy (i.e., which and how many production microservices are cloned) plays a fundamental role in taking advantage of this behavior and maximizing the chance of attacker interaction. An effective decoy placement scheme should consider the modification introduced by decoys on the original microservice deployment and tailor the cloning mechanism according to the altered perception of the attacker about the target environment. 

Although research on decoy placement schemes has addressed some of the challenges related to this allocation problem, it has limitations. Most of the proposed strategies assume that decoys can be located near the organization's network perimeter and thus cannot handle threats that elude the first line of defenses or insider attacks propagating within the production environment \cite{zhang2021three}. Additionally, these schemes often consider a fixed number of decoys that need to be allocated on a predetermined number of system assets. As a consequence, the scaling efficiency as well as the resource consumption of these techniques is not addressed and cannot be extended to support large-scale microservice deployments where every microservice can be potentially cloned as decoy. Finally, many of these approaches rely on game-theoretical models to determine a suitable decoy allocation strategy \cite{zhu2019game}\cite{anwar2020game}. 
Such resolution methodology cannot properly capture neither the heterogeneous nature of the resources (e.g. CPU, memory, disk storage) nor the constraint on the maximum availability of those resources in the considered physical infrastructure. 

Motivated by such research gap, this paper presents a novel optimization-based decoy allocation scheme for cloud-native microservice architectures. In detail, our work provides the following contributions:
\begin{itemize}
            \item we design a metric to evaluate the decoy effectiveness in luring attacks according to the attack graph structure. The latter models the admissible lateral movements of an attacker between microservices. The proposed formulation accounts for the topology modifications in the attack graph introduced by the decoy deployment in order to fully characterize the deceptive capability of each new decoy placement.
            \item we formulate an integer non-linear optimization problem leveraging the proposed metric in order to compute the optimal decoy placement strategy. By including the impact of the allocated decoys on the attack graph within the objective function formulation, our solution maximizes the number of attack paths intercepted by the decoys according to the availability of computational resources.
            \item we design an heuristic scheme to overcome the computational complexity of the optimal formulation. The proposed algorithm approximates the original objective function by leveraging an iterative decoy placement strategy which greatly reduces complexity while ensuring a modest performance loss compared to the optimal solution.
            \item we evaluate optimal and heuristic solutions against \hl{state-of-the-art and baseline schemes}  using several metrics to assess the effectiveness of our resource-aware decoy placement strategy. Our schemes outperformed the benchmark ones in terms of deceptive capability performance, while requiring approximately the same number of decoys.
\end{itemize}
The remainder of the paper is structured as follows. In Section II, we discuss the related work. In Section III, we describe the system model and each component. In Section IV, we propose an optimization-based decoy allocation scheme as well as a heuristic algorithm to approximate the optimal solution.  In Section V, we assess the performance of the proposed schemes. Finally, we draw the conclusion in Section VI.

\section{Related work}
Modern cyber deception has evolved towards the design of flexible and adaptive deception techniques tailored to the considered defense scenario in order to maximize the deception effectiveness and reduce the management cost complexity. A more detailed discussion of cyber deception principles and related challenges can be found in \cite{wang2018cyber} and \cite{zhu2021survey}. In the context of microservice architectures, deception strategies need lightweight and flexible decoy implementations to minimize the impact on the computational resources \cite{gotz2018challenges}. Moreover, the distributed nature of microservice-based applications requires the design of fine-granularity decoy allocation schemes to efficiently cover the wide attack surface characterizing this technology  \cite{yarygina2018overcoming}.
Following these observations, we discuss the work on decoy/honeypot allocation schemes by focusing the analysis on the related resource efficiency and deceptive effectiveness of the decoy placement. 

The authors of \cite{osman2019sandnet} develop a sandbox network that misdirects attacks from the production network towards deceptive containers with exposed vulnerabilities that are deployed in a isolated and monitored virtual environment. Although their approach allows to trap the attacker within a confined environment separated from the real assets, both the management cost and the resource consumption performance of the decoy are not addressed. We instead limit the orchestration complexity by directly disseminating deceptive microservices replicas within the production infrastructure according to a given resource budget.

In \cite{li2021cyber}, the authors propose a fingerprint anonymity technique to slow and increase the difficulty of the attacker reconnaissance activities by generating container replicas having a fake set of labels attributes. Differently, our decoy allocation scheme deceives the attacker lateral movements between microservices by employing fully-interactive microservice replicas that are clone of the legitimate ones.

The authors of \cite{li2022optimal} and \cite{li2022defensive} design a deep reinforcement learning agents that place fake microservices replicas in order to maximize their capability to lure attacks directed toward legitimate microservices and to conceals assets, respectively. Similarly, \cite{anwar2022honeypot} proposes a two-phase honeypot allocation algorithm combining game theory and reinforcement learning techniques in order to model and dynamically adapt the honeypot allocation according to the attacker activity.
Beside the omission of the decoy resource consumption, the main limitation of \cite{li2022optimal}\cite{li2022defensive}\cite{anwar2022honeypot} is the scaling efficiency of the proposed schemes that mostly derived from machine learning approaches. Specifically, these algorithms suffers from a high computational complexity as the training procedure requires many iterations to converge toward a stable allocation policy when the number of possible decoy allocation configurations increases. Instead, we tackle the design of a resource-aware decoy allocation scheme using an optimization-based approach that allows to extend the solution for a high number active microservices by leveraging a low-complexity heuristic.

The authors of \cite{liu2021deception} design a honey-credential deployment strategy that allocates fake account/password  to maximize the probability to lure attacks. Although their approach ensures a very low resource usage as the decoy are simple text files, our solution still ensures a manageable resource consumption overhead and, at the same time,  high-interaction decoys that can provide valuable information on the attacker TTPs when violated.

In \cite{anwar2022cyber}, the authors formulate a honeypot placement scheme given a limited budged of resources. In particular, the resulting allocation policy leverages a game-theoretical approach and incentivizes the placement of decoys on the most valuable network assets. However, the considered resources are homogeneous whereas we consider heterogeneous computational resource such CPU cycles and RAM volume needed to run each decoy. Furthermore, the placement strategy follows a sidecar approach which aims at defending vulnerable network entities by exclusively allocating decoys on those assets. Differently, we aggregate the exploit difficulty of the various microservices in order to place decoys on critical microservices whose violation would allow the attacker to spread more effectively among the infrastructure. 

The authors of \cite{jin2019dseom} \hl{propose a scheme that first identifies the most vulnerable assets in a cloud environment based on the number of attack paths traversing them, then it uses such information to deploy suitable moving target defense strategies. 
However, such an approach neglects the modification in the number of attack paths affecting the legitimate microservice due to allocation of the deceptive microservices. Differently, we provide an analytical formulation that updates the number of attack paths traversing the legitimate microservices for any decoy placement configuration and we employ this metric to design an optimization problem maximizing the number of attack paths that are intercepted by the allocated decoys.}

Finally, the authors of \cite{ma2023optimal} and \cite{amin2019online}  propose statistic approaches based on probabilistic attack graphs and partially observable Markov decision process, respectively, in order to provide a decoy allocation strategy according to the attacker uncertain behavior. However, they assume that the decoy placement is restricted to specific locations within the infrastructure and neglect the impact of the decoys on the attack graph topology, thus limiting the approaches flexibility. Differently, we assume that any microservice can be replicated as decoy and we design our decoy placement scheme based on the worst-case attack path scenario.

\hl{To summarize, unlike the previously discussed work, our cyber deception scheme provides a lightweight and scalable decoy placement strategy that maximizes the chance of the attacker interaction according to a predetermined amount of computational resources. Our solution can be integrated within the production environment of cloud-native applications thanks to the flexibility offered by container-based technologies. In detail, we generate high-interaction decoys by cloning active microservices of an application and we deploy them as legitimate microservices that effectively blend within the defender environment.} 

\section{System model}
\begin{figure}[t]
    \centering
    \includegraphics[width=0.95\linewidth]{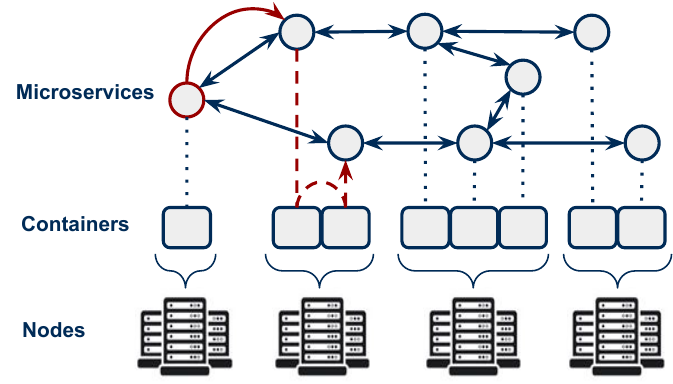}
    \caption{Overview of the considered system model. 
    Microservices can be compromised by the attacker using remote code execution techniques (solid red arrow) or container escape techniques (dashed red arrow).}
    \label{fig:architecture}
\end{figure}

We discuss the system model that we considered in order to design the decoy allocation strategy.  We provide a comprehensive representation of the main system features in Fig.\ref{fig:architecture}. In the following subsections, we describe the various components identified as microservice architecture model, decoy configuration and placement model and threat model. In Table \ref{glossary}, we resume the symbol notation.

\begin{table}[t]
	\centering
	\scriptsize
	\caption{Glossary of symbols.}
	\vspace{-2mm}
	\label{tab:notations}
	\begin{tabular}{c}
		\textit{Sets}
	\end{tabular}
	\\
	\renewcommand{\arraystretch}{1.2}
	\begin{tabular}{|p{1.2cm}|p{6.8cm}|}
		\hline
		\textit{N} & Set of computing nodes \\
		\hline
		\textit{M} & Set of production microservices \\
		\hline
		\textit{D} & Set of allocated decoys \\
		\hline
		\textit{V} & Set of vertices in the attack graph \\
		\hline
		\textit{E} & Set of directed edges in the attack graph \\
		\hline
	\end{tabular}
	\\
	\vspace{1mm}
	\begin{tabular}{c}
		\textit{Parameters}
	\end{tabular}
	\\
	\renewcommand{\arraystretch}{1.4}
	\begin{tabular}{|p{1.2cm}|p{6.8cm}|}
		\hline
		\textit{$d_m^{(j)}$} & $d_m^{(j)}\in D$ is the j-th decoy replica of microservice $m\in M$\\
		\hline
		\textit{$v_i$} & $v_i\in V$ vertex in the attack graph\\
		\hline
		\textit{$e_{ij}$} & {$e_{ij}=(v_i,v_j)$} edge between vertices $v_i$ and $v_j$ in the attack graph\\
		\hline
		\textit{$C^{cpu}_{n}$} & Maximum CPU capacity of node $n\in N$\\
		\hline
		\textit{$C^{ram}_{n}$} & Maximum RAM capacity of node $n\in N$\\
		\hline
		\textit{$r^{cpu}_{m,n}$} & CPU resources requested by microservice $m \in M$ on node $n\in N$\\
		\hline
		\textit{$r^{ram}_{m,n}$} & RAM resources requested by microservice $m \in M$ on node $n\in N$\\
		\hline
		\textit{$\Delta C^{(.)}_n$} & Amount of available resources (CPU or RAM) on node $n\in N$\\
		\hline
		\textit{$\delta$} & \bc{Decoy resource ratio, $\delta \in [0,1]$ }\\
		\hline
		\textit{$EM_{\nu}$} & Exploitability score of vulnerability $\nu$ \\
		\hline
		\textit{$ECM_{\nu}$} & Likelihood score of vulnerability $\nu$ being attacked \\
		\hline
		\textit{$AP_{(v_s,v_t)}$} & Attack path between vertices $v_s$ and $v_t$ \\
		\hline
		\textit{$DAP_{(v_s,v_t)}$} & Deceptive attack path between vertices $v_s$ and $v_t$ \\
		\hline
        \textit{$\sigma^{(v_i)}_{AP_{(v_s,v_t)}}$} & Binary function such that $\sigma^{(v_i)}_{AP_{(v_s,v_t)}}=1$     iff $v_i\in AP_{(v_s,v_t)}$\\
        \hline
		\textit{$\bar{\sigma}^{(v_i)}_{AP_{(v_s,v_t)}}$} & Binary function such that $\bar{\sigma}^{(v_i)}_{AP_{(v_s,v_t)}}=\sigma^{(v_i)}_{AP_{(v_s,v_t)}}$ iff $D = \emptyset$\\
		\hline
		\textit{$\bar{b}(v_i)$} & Betweenness centrality value of vertex $v_i\in V$\\
		\hline
		
	\end{tabular}
	\\
	\vspace{1mm}
	\begin{tabular}{c}
		\textit{Decision variables}
	\end{tabular}
	\renewcommand{\arraystretch}{1.2}
	\begin{tabular}{|p{1.2cm}|p{6.8cm}|}
		\hline
		\textit{$x_m$} & Number of decoys cloning microservice $m\in M$ \\
		\hline
		
	\end{tabular}
	\\
\label{glossary}
\end{table}

\subsection{Microservice architecture model}

We consider an organization running microservice-based applications on the cloud. Every microservice is virtualized and deployed as a container. We indicate as $M$ the set of production microservices, whereas we indicate as $N$ the set of computing nodes. Each node has a fixed capacity in terms of processing and memory capabilities denoted as $C_{n}^{cpu}$ and $C_{n}^{ram}$, respectively. 
 We analytically represent the deployment of each microservice on a specific node by defining variables $r_{m,n}^{(cpu)}$ and $r_{m,n}^{(ram)}$, which express the amount of CPU and RAM resources requested by microservice $m$ to run on node $n$, respectively. 
In particular, if microservice $m$ is assigned to node $n$, then $r_{m,n}^{(cpu)}=\bar{r}_{m}^{(cpu)}$ and $r_{m,n}^{(ram)} = \bar{r}_{m}^{(ram)}$, where $\bar{r}_{m}^{(cpu)}$ and $\bar{r}_{m}^{(ram)}$ represent the specific values of CPU and RAM resources. Otherwise, if microservice $m$ is not assigned to node $n$, then $r_{m,n}^{cpu} = 0$  and $r_{m,n}^{ram} = 0$. Based on this notation, we define the resource deployment matrix as 

\begin{multline}
   \mathbf{R}^{(.)} = [
        r_{m,n}^{(.)} : \sum_{m \in M} r_{m,n}^{(.)} \leq C_{n}^{(.)} \quad \land \\
        \sum_{n \in N} r_{m,n}^{(.)} = \bar{r}_{m}^{(.)}, \forall m \in M, \forall n \in N],
        \label{eq:deployment_matrix} 
\end{multline}
where the $^{(.)}$ notation indicates that the above formulation holds both for CPU and RAM resources. 
The first condition in \eqref{eq:deployment_matrix} ensures that the total number of resources requested by each microservice cannot exceed the node capacity. Similarly, the second condition guarantees that the microservice deployment is feasible by uniquely assigning a single node to each microservice. 

\subsection{Decoy configuration and placement model}
\begin{figure}[t]
     \centering
     \begin{subfigure}{0.4\columnwidth}
         \centering
         \includegraphics[width=0.8\linewidth]{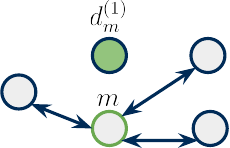}
         \subcaption{Microservice $m$ is cloned as decoy $d_{m}^{(1)}$.}
         \label{fig:decoy_placement_no_decoy}
     \end{subfigure}
     \hspace{1cm}
     \begin{subfigure}{0.4\columnwidth}
         \centering
         \includegraphics[width=0.8\linewidth]{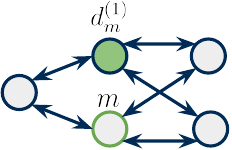}
         \subcaption{The interfaces of decoy $d_{m}^{(1)}$ are configured as $m$.}
         \label{fig:decoy_placement_one_decoy}
     \end{subfigure}
     \caption{Example of deceptive microservice allocation within a  microservice architecture.}
     \label{fig:decoy_deployment}
\end{figure}

Each decoy is created by cloning the image of production microservices as well as the related interfaces towards adjacent communicating microservices.  In Fig. \ref{fig:decoy_deployment}, we show an example of decoy placement that can be easily extended to any microservice deployment topology. Furthermore, every decoy is configured to provide the following features: 
\begin{itemize}

\item{\textit{Attack detection reliability}: any data traffic intercepted by a decoy is considered malicious since legal data traffic is exclusively forwarded to production microservices. For example, this behavior could be implemented by defining specific network policies with Kubernetes \cite{budigiri2021network}. This tool makes it possible to steer the application traffic flow among legitimate microservices while preserving their interfaces towards the decoys. Based on this configuration, the attacker can still interact with the decoys as the latter are only isolated from the application traffic and not from networking reachability.}

\item{\textit{High interactivity}: each decoy reacts to the attacker input like a production microservice. However, any data extracted by the attacker is fake and its content is configured to resemble the structure of production data (for example, a database filled with fake entries).}

\item{\textit{TTPs monitoring}: each decoy implementation is augmented by monitoring functionalities that allow the defender to gather cyber threat intelligence information on the attacker TTPs employed to violate the decoy.}
\end{itemize}
The advantage of the proposed decoy configuration is twofold. It reduces the likelihood that the attacker can spot the deception mechanism as decoys are embedded within the microservice deployment topology like production microservices. It limits the decoy management complexity by duplicating microservices that are currently running. 

Instead of cloning the up-to-date implementation of a production microservice, a deceptive microservice could employ a legacy implementation version characterized by known and non-patched vulnerabilities. On one hand, this design choice increases the decoy luring capability. On the other hand, beside possible compatibility issues with the current microservice deployment, it enlarges the attack surface of the infrastructure by intentionally introducing additional weaknesses that offer new TTPs to the attacker. For this reason, we do not consider such a decoy configuration option and we plan to better investigate such tradeoff in a future work to improve the proposed decoy allocation strategy.


We indicate the set of allocated decoys as $D = \{d^{(j)}_m, m \in M, j \in \mathbb{N}_{>0}\}$, where $d^{(j)}_m$ refers to the $j$-th decoy replicating microservice $m$. Due to the resource limitation, a microservice can be replicated as decoy only if the resources reserved to the deception mechanism satisfies the same CPU and RAM resource request associated to the legitimate microservice. Therefore, assuming that microservice $m$ is allocated on node $n$, we express a feasible decoy allocation as 
\begin{multline}
    \mathbf{x} = [x_m: x_m  \bar{r}_{m}^{(.)} \leq \delta \cdot \Delta C_{n}^{(.)},\forall n \in N],
    \label{eq:decoy_allocation} 
\end{multline}
where $x_m = \{d^{(j)}_m\}$ indicates the total number of decoys cloning microservice $m$, $\delta \in  [0,1]$ denotes the \bc{decoy resource ratio} and indicates the fraction of available resources that are reserved for the decoy allocation, $\Delta C_{n}^{(.)}$ is the amount of unused resources in each node given the microservice resource deployment  $\mathbf{R}^{(.)}$. The condition in \eqref{eq:decoy_allocation} ensures that the total resource request of decoys cloning a specific microservice can be accommodated by the available resources in the node.
We assume that $\delta$ \hl{is fixed and its value is selected by the organization in order to balance the security level and the resource overhead achieved by the deception mechanism.} Specifically, the higher is the value of $\delta$, \hl{the higher is the number of deployable decoys as more resources are dedicated for the deception mechanism which intuitively allows to lure and intercept more attacks.} By computing the decoy allocation strategy according to a configurable portion of the available resources, we mitigate potential impacts on the production microservices performance due to resource contention issues.

\subsection{Threat model}

\hl{We consider an attack scenario where a malicious user (the attacker) has eluded the security mechanisms of an organization (the defender), such as IDSs and firewalls, and gained access to his cloud services by compromising one of the deployed microservices. 
The initial access can be accomplished by means of phishing mails, stolen login credential or exploits of any internet-facing service (e.g. web servers) hosted by the targeted organization. Upon entering the defender environment, the attacker gathers information about the configurations of deployed microservices. For example, he can inspect the environment variables associated to each container running in a Kubernetes cluster to discover IP address, port number and service name of other microservices within the deployment \cite{sayfan2017mastering}. This information allows him to remain undetected as well as to plan his strategy according to deployment vulnerabilities.
Due to the loose coupling between microservices as well as the distributed nature of cloud-native  deployments, it is reasonable to assume that the attacker can further penetrate the defender environment by moving laterally among the microservices of the violated application } \cite{mateus2021security}. \hl{However, we assume he is unaware of the deception mechanism, thus he can compromise any legitimate microservice and/or decoy in order to reach his goal.}
The attacker lateral movements can be accomplished by means of two techniques:
\begin{itemize}
\item \textit{Remote Code Execution} (RCE): the attacker can violate a microservice by injecting malicious code throughout its interfaces \cite{alshamrani2019survey}. The compromised microservice is used as foothold to violate other communicating microservices. 
\item \textit{Container Escape}: 
\hl{the attacker exploits any non-root privilege configuration available in each container to interact with other containers that do not belong to the same name-space} \cite{sultan2019container}.
In other words, we assume that the attacker can break the microservices logical isolation in order to compromise any microservice deployed on the same node.
\end{itemize}
We assume that each decoy is deployed on the same node of the original cloned microservice. From a security perspective, although this assumption restricts the possible number of decoy placement strategies, this constraint negates the addition of new container escape threats that would be generated by instead allocating a decoy on a different node. Moreover, regardless of the employed technique, we assume that the attacker cannot accomplish a privilege escalation to gain access to the whole cluster (i.e., the full set of nodes running the containerized microservices). This attack scenario would bypass any deployed defense mechanism and it is mostly due to a poor configuration of the system privilege levels performed by the defender, hence it is beyond the deception scope. 

We represent the attacker lateral movements in the considered system using graph theory in order to better tailor the design of an effective decoy allocation strategy \cite{milani2020harnessing}. This mathematical tool can efficiently model the attacker lateral movements among microservices and decoys by abstracting the attacker techniques in the form of a sequence of graph vertices and edges. 
In detail, we introduce the \textit{Attack Graph} (AG) associated to the current microservice deployment configuration as a directed acyclic graph $G=(V,E,f)$ where:
\begin{itemize}
\item $V = \{v_i \in M \cup D\}$ is the set of vertices in the graph corresponding to the active microservices and decoys.
\item $E = \{(v_i,v_j) : v_i \prec v_j \quad \land  \quad v_i, v_j \neq v_i \in V \times V\}$ is the set of directed edges identified by the ordered vertex pairs $e_{ij} = (v_i,v_j)$ that represent the microservices downstream call flow (in other words, we consider that microservice $i$ requests a task to microservice $j$). More specifically, the vertex pair $(v_i,v_j)$ is connected by the directional edge $e$ if an attacker can move laterally from microservice $i$ to microservice $j$ by either leveraging RCE or container-escape techniques. Note that in the latter scenario, microservices $i$ and $j$ must be allocated on the same computing node.  
\item $f: \{E \rightarrow w\}$ is a function assigning a weight $w \in \mathbb{R}^+$ to each directed edge $e_{ij}$ expressing the microservice vulnerability level. Note that the same reasoning applies to decoys.
In order to formalize this concept, \bc{we follow an approach similar to the one adopted by the authors of \cite{jin2019dseom} that employed metrics defined in Common Vulnerability Scoring System (CVSS) to quantify the exploiting difficulty of software applications. In particular, they used the \textit{Exploitability Metrics} (EM) and \textit{Exploit Code Maturity} (ECM) indicators  \cite{cvss2019common}. The former measures the difficulty of exploiting a vulnerability of an application and depends on different factors such as the privileges required by the adversary and the level of interactivity required from the user during the exploit process. Differently, the latter indicates the
likelihood of that vulnerability being attacked and usually depends on the complexity of the techniques employed by the adversary to finalize the exploit process}. Leveraging these metrics, we compute the weight for each edge $e_{ij}$ as 
\begin{equation}
    f(e_{ij}) = \frac{\sum\limits_{\nu \in \mathcal{V}_j} EM^{\bc{(j)}}_{\nu} \cdot ECM^{\bc{(j)}}_{\nu}}{\sum\limits_{\nu \in \mathcal{V}_j} ECM^{\bc{(j)}}_{\nu}} \quad \forall i \in V
    \label{eq:weight}
\end{equation}
where $\mathcal{V}_j$ is the set of vulnerabilities associated to microservice $j \in M$. Note that $\mathcal{V}_j$ also includes the virtualization vulnerability when microservice $i$ is allocated on the same node as microservice $j$.
\bc{In practice, each vertex defines the weight $w$ of its in-ward edges as the average value of the $ECM^{\bc{(j)}}_{\nu}$ scores, each one weighted by the corresponding $EM^{\bc{(j)}}_{\nu}$, associated to microservice $j$. We employ low numerical values of $ECM^{\bc{(j)}}_{\nu}$ and $EM^{\bc{(j)}}_{\nu}$ to describe microservices with vulnerabilities that can be easily exploitable. As a result, low values of edge weights  indicate less secure microservices.}
\end{itemize}

We assume that the attacker reaches his target by violating the sequence of microservices providing the ``least impedance" path in terms of exploitability complexity given his entry-point.
Although this assumption disregards alternative attacker profiles, such as those who engage in stealthy lateral movements to minimize detection or maximize damage to the defender's environment, we corroborate the effectiveness of our approach by demonstrating its resilience to various attacker behaviors in the results section. We quantify such attacker behavior as the shortest path in AG between a source vertex $v_s \in V$, that identifies the attacker entry-point, and a target vertex $v_t \in V$, that identifies the target microservice whose violation makes it possible for the attacker to access some organization assets. Formally, we define the $Attack$ $Path$ (AP) between $v_s$ and $v_t$ as the sequence of vertices
\begin{equation}
        AP(v_s,v_t) = [(v_s,\dotsc,v_i,\dotsc,v_t) : (v_i,v_{i+1}) \in E, \forall v_i \in V]
\end{equation}
that minimizes the total weight $\sum_{e \in AP(v_s,v_t)} f(e)$. Intuitively, each $AP(v_s,v_t)$ 
represents the worst-case attack scenario where a malicious user employs the most efficient sequence of techniques to penetrate the system.
Moreover, if an AP contains at least one decoy between $v_s$ and $v_t$, we denote this path as a $Deceptive$ $Attack$ $Path$ (DAP), i.e.
\begin{equation}
        DAP(v_s,v_t) = [AP(v_s,v_t) : AP(v_s,v_t) \cap D \neq \{\phi\}]
        \label{eq:deceptive_ap}
\end{equation}
The amount of DAPs in AG measures the deception quality of the decoy placement configuration since the higher is this value, the higher is the number of attacks that are likely to be intercepted by the decoys. From the above definitions, we remark that a DAP is also an AP, hence we can generally refer to any path in the AG as an AP if needed. 

\section{Problem formulation}

\begin{figure}[t]
     \centering
     \begin{subfigure}{0.5\textwidth}
         \centering
         \includegraphics[width=0.8\linewidth]{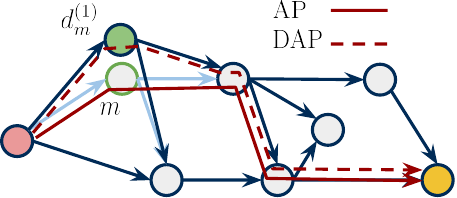}
          \subcaption{DAP generated by the allocation of decoy $d_{m}^{(1)}$.}
         \label{fig:DAP_d1}
     \end{subfigure}
     \par\bigskip
     \begin{subfigure}{0.5\textwidth}
         \centering
         \includegraphics[width=0.8\linewidth]{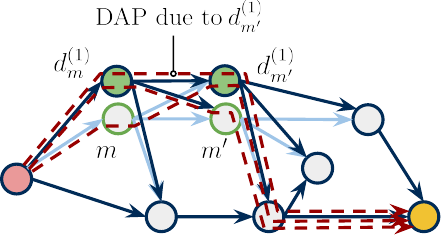}
         \subcaption{DAPs generated by the allocation of decoys $d_{m}^{(1)}$ and $d_{m'}^{(1)}$.}
         \label{fig:DAP_d2}
     \end{subfigure}
     \caption{Example of DAPs generated by a different number of decoys along a specific AP in AG. The red and yellow microservices represent the AP source and target, respectively.}
     \label{fig:DAP_generation}
\end{figure}

Assuming that the defender is unaware of the attacker entry point and targets, an effective decoy allocation maximizes the number of generated DAPs between any vertex pair in AG according to the resource availability.  
We illustrate this intuition by analyzing the impact of the decoy allocation on the AG topology in Fig. \ref{fig:DAP_generation}. The latter depicts an example of DAP generation in a simple scenario but it can be easily extended for an arbitrary number of decoys and APs. Each decoy placement modifies the AG structure by adding a new “deceptive" vertex having the same features as the one corresponding to the replicated microservice. As a consequence, a decoy generates a number of DAPs equals to the number of APs traversing the cloned microservice (Fig. \ref{fig:DAP_d1}). Note that those APs may also include DAPs that are introduced by other allocated decoys as highlighted in Fig. \ref{fig:DAP_d2}, where decoy $d_{m}^{(1)}$ now generates an additional DAP due to the allocation of $d_{m'}^{(1)}$ on the same AP (the same argument applies for the additional DAP traversing $d_{m'}^{(1)}$ due to  $d_{m}^{(1)}$).

Following these observations, the allocation of decoys on microservices sharing multiple APs can produce a high number of DAPs and thus increases the decoys likelihood to misdirect and intercept possible attacks. In other words, using an analogy,  decoys can be considered as ``magnets'' that attract APs to their location, hence an effective decoy placement is the one that maximizes the total number of intercepted APs to maximize the number of generated DAPs.

We formalize the dynamic of decoy allocation by using the concept of \textit{Betweenness} \textit{Centrality} (BC) of a vertex. This metric measures the number of shortest paths traversing each graph vertex and offers an analytic approach to tackle the decoy allocation problem. The higher this value is, the more “central" is the related vertex compared to the surrounding vertices since its location allows to reach multiple destinations with minimum cost. \hl{Based on this definition, we can consider the BC as a security metric indicating the likelihood that a microservice may eventually interact with an attacker. Consequently, a suitable decoy allocation strategy should select those microservices with high BC values to increase the chance of the decoys to intercept possible attacks.}

Formally, we define $\sigma^{(v_i)}_{AP(v_s,v_t)}$ as the binary indicator function that expresses whether the AP from $v_s$ to $v_t$ contains the vertex $v_i$. More specifically, $\sigma^{(v_i)}_{AP(v_s,v_t)} = 1$ if $v_i \in AP(v_s,v_t)$ and $\sigma^{(v_i)}_{AP(v_s,v_t)} = 0$ otherwise. Similarly, we define $\sigma_{AP(v_s,v_t)}$ as the variable indicating the number of APs from $v_s$ to $v_t$. According to this notation, we compute the BC of a vertex $v_i \in V$ given the number of APs in $AG$ as
\begin{equation}
    b(v_i)  = \sum_{v_s \neq v_i \in V} \sum_{v_t \neq v_s \in V} \frac{\sigma^{(v_i)}_{AP(v_s,v_t)}}{\sigma_{AP(v_s,v_t)}}.
    \label{eq:betweenness_formulation}
\end{equation}
The denominator of \eqref{eq:betweenness_formulation} can be considered as a normalization factor balancing the BC gain contribution of the multiple equal-cost DAPs generated by the decoys allocated on the same microservice.

\subsection{Optimal decoy allocation}

In practice, $b(v_i)$ allows to formally quantify the deception effectiveness of each decoy according to its placement. However, the formulation \eqref{eq:betweenness_formulation} cannot be employed as optimization objective since it is not expressed as a function of the number allocated decoys, which is indeed the optimization variable. In general, the computation of the BC of a vertex given an arbitrary number of graph modifications (e.g. vertex addition/removal and edge addition/removal) is a very challenging task since the corresponding shortest paths modifications can hardly be predicted \cite{bader2007approximating}. 

However, by leveraging the AG topology structure as well as the decoy allocation dynamic, we can overcome this issue and propose an alternative formulation of $b(v_i)$ fulfilling the aforementioned requirement. 
In detail, we employ the initial number of APs going through each vertex in the original AG without decoys as a basis to update the BC value of any vertex when a decoy allocation is computed. As a matter of fact, the decoy allocation, which corresponds to a vertex addition operation, preserves the acyclic property of AG since the new introduced edges have the same direction of the in-ward and out-ward edges of the cloned microservice (see Fig. \ref{fig:DAP_generation}). 

This particular feature can be used to exactly predict the number of DAPs generated by any decoy placement configuration.
Formally, we compute the BC of a vertex $v_i \in V$ given a feasible decoy allocation $\mathbf{x}$ as
\begin{equation}
    \bar{b}(v_i) =  \sum_{v_s \neq v_i \in V} \sum_{v_t \neq v_s \in V} \bar{\sigma}^{(v_i)}_{AP(v_s,v_t)} \cdot \frac{(1+x_s) \cdot (1+x_t)}{(1+x_i)},
    \label{eq:betwenness_formulation_decoy}
\end{equation}
where $x_s,x_t,x_i$ represent the number of decoys allocated on microservices corresponding to vertices $v_s,v_t,v_i$, respectively, and $\bar{\sigma}^{(v_i)}_{AP(v_s,v_t)} = 1$ indicates if there exists an AP going through vertex $v_i \in M$ when no decoys are allocated, $\bar{\sigma}^{(v_i)}_{AP(v_s,v_t)} = 0$ otherwise. The numerator of addend in \eqref{eq:betwenness_formulation_decoy} provides the total number of DAPs traversing vertex $v_i$ that are generated from the vertex pairs $(v_s,v_t)$ having $x_s$ and $x_t$ decoys allocated, respectively.
 Note that DAPs generated by vertex pairs not containing $v_i$ are excluded from the computation since $\bar{\sigma}^{(v_i)}_{AP(v_s,v_t)} = 0$ by construction.
In practice, the total number of DAPs is computed by summing up the individual contribution of the possible vertex pairs formed between the sets $\{v_s,v_{d^{(1)}_s},...,v_{d^{(x_s)}_s}\}$ and  $\{v_t,v_{d^{(1)}_t},...,v_{d^{(x_t)}_t}\}$. 
Each vertex pair enumerates the DAP between the corresponding source and target vertices as $\sigma^{(v_i)}_{AP(v_j,v_k)} = 1$ where $ j = \{s, d^{(1)}_s, \dotsc , d^{(x_s)}_s\}$ and $k = \{t, d^{(1)}_t, \dotsc , d^{(x_t)}_t\}$.
The denominator of addend in \eqref{eq:betwenness_formulation_decoy} accounts for the fact that any DAP generated by each vertex pair is also replicated by decoys allocated on microservice $i$ if $x_i > 0$, hence it acts as normalization factor like the denominator of \eqref{eq:betweenness_formulation}. 

Leveraging this alternative BC formulation, we compute the decoy allocation maximizing the total number of DAPs given the microservice resource deployment configuration expressed by $\mathbf{R}^{(.)}$ as
\begin{gather}
       \max_{\mathbf{x}} \sum_{m \in M} x_{m} \cdot \bar{b}(v_m)
       \label{eq:obj_func}
\end{gather} 
subject to
\begin{equation}
   \sum_{m \in M} x_{m}r_{m,n}^{(cpu)} \leq \delta \cdot \Delta C_{n}^{(cpu)} \quad \forall n \in N
   \label{eq:const_cpu}
\end{equation}
\begin{equation}
   \sum_{m \in M} x_{m} r_{m,n}^{(ram)} \leq \delta \cdot \Delta C_{n}^{(ram)} \quad \forall n \in N
   \label{eq:const_ram}
\end{equation}
\begin{equation}
   x_{m} \cdot (r_{m,n}^{(cpu)} - \bar{r}_{m}^{(cpu)}) \geq 0 \quad \forall m \in M, \forall n \in N
   \label{eq:const_single_node_cpu}
\end{equation}
\begin{equation}
   x_{m} \cdot (r_{m,n}^{(ram)} - \bar{r}_{m}^{(ram)}) \geq 0 \quad \forall m \in M, \forall n \in N
   \label{eq:const_single_node_ram}
\end{equation}
\begin{equation}
   x_m \in \mathbb{N},  \quad \forall  m \in M
   \label{eq:const_x}
\end{equation}


Each addend in \eqref{eq:obj_func} provides the number of DAPs generated by the $x_m$ decoys replicating microservice $m$. Constraints \eqref{eq:const_cpu} and \eqref{eq:const_ram} ensure that the allocated decoys do not exceed the dedicated CPU and RAM resources defined by parameter $\delta$, respectively. Constraints \eqref{eq:const_single_node_cpu} and \eqref{eq:const_single_node_ram} enforce that each decoy must be allocated on the same computing node of the cloned microservice. 
Finally, constraint \eqref{eq:const_x} expresses the integer nature of the considered allocation problem.

The proposed optimization problem is characterized by a non-linear objective function as it can be easily observed by the definition of $\bar{b}(v_i)$ in \eqref{eq:betwenness_formulation_decoy} which makes the solution computation very demanding. Moreover, the integrity constraint further exacerbates this issue as it renders the problem combinatorial. Intuitively, \eqref{eq:obj_func}-\eqref{eq:const_x} can be seen as a Knapsack problem with variable coefficients $\bar{b}(v_i)$ and thus its complexity is NP-hard \cite{suzuki1978generalized}. For this reason, this problem formulation is unpractical for large-scale microservice deployments that are often reconfigured as this would also require the recomputation of the decoy allocation. 
To overcome this limitation, we approximate the optimal allocation by designing a heuristic decoy allocation scheme to reduce the solution computational complexity.

\subsection{Heuristic decoy allocation}

The proposed heuristic consists of a greedy algorithm that prioritizes the allocation of decoys on microservices that require a low amount of resources and, at the same time, that are traversed by a high number of APs (in other words, it employs the BC values associated to each microservice in the original AG). In detail, the algorithm allocates one decoy at every iteration and update the DAPs generation according to the previously allocated decoys. This procedure makes it possible to progressively enumerate every DAP introduced by the decoys and thus can help approximating the optimal allocation. We present the pseudo-code for this scheme in Algorithm \ref{alg:heuristic}. 

\begin{algorithm}[t] 
\caption{Heuristic decoy allocation}\label{alg:heuristic}
\begin{algorithmic}[1]
\State \textbf{Input:} AG, $\bar{\sigma}$, $b$, $ \mathbf{R}^{(.)}$, $\Delta C^{(.)}$
\State \textbf{Output:} $\mathbf{x}$
\State Initialize $x_m = 0$ for each $m \in M$ 
\State Initialize the priority queue $Q$ to sort microservices
\State Compute $\hat{d}_m = \lfloor \delta \Delta C^{(.)}_n / \bar{r}^{(.)}_{m} \rfloor$ for each $m \in M, n \in N$ \label{algol:initial_cost}
\For{each m $\in$ $M$}
    \If{$\hat{d}_m$ $\geq$ 1}
         \State $p$ $\leftarrow$ -$b(v_m)$ $\cdot$ $\hat{d}_m$ \label{algol:initial_priority}
        \State Insert $m$ into $Q$ with priority $p$ \label{algol:first_queue_insertion}
    \EndIf
\EndFor
\While{$Q$ is not empty} 
    \State Extract the first $m$ from $Q$ \label{algol:extract_from_queue}
    \State Allocate a decoy on $m$ and update the AG topology 
    \State Update $x_m \leftarrow x_m + 1$ \label{algol:deploy_decoy}
    \State Update resource availability $\delta \Delta C^{(.)}_n \leftarrow \delta \Delta C^{(.)}_n - \bar{r}^{(.)}_{m}$\label{algol:update_resources}
    \For{each $i$ $\in$ $M$} \label{algol:start_sideffect}
        \For{each $t$ $\in$ $M$}
            \State $b(v_i)$ $\leftarrow$ $b(v_i)$ + $( \bar{\sigma}^{(v_i)}_{AP(v_m,v_t)} + \bar{\sigma}^{(v_i)}_{AP(v_t,v_m)} )$ $\cdot$ \phantom{xxxxxxxxxxxxxxxx} ($x_t + 1$) 
            \label{algol:update_betweenness_1}
            \State $\theta^{(v_i)}_{v_t} \leftarrow \theta^{(v_i)}_{v_t}$ + $( \bar{\sigma}^{(v_i)}_{AP(v_t,v_m)} + \bar{\sigma}^{(v_i)}_{AP(v_m,v_t)} )$ $\cdot$ \phantom{xxxxxxxxxxxxxxxx} $(x_m + 1)$ 
            \label{algol:update_gains_1}
        \EndFor
    \EndFor \label{algol:end_sideffect}
    \State Clear $Q$ \label{algol:queue_clear}
    \State Recompute $\hat{d}_m$ for each microservice \label{algol:update_ratios}
    \For{each $i$ $\in$ $M$}
        \State $\Delta b(v_i)$ $\leftarrow$ $b(v_i)$ $\cdot$ ($x_i$ + 1) / ($x_i$ + 2) 
        \label{algol:start_compute_reward}
        \For{each $t$ $\in$ $M$}
            \State $\Delta b(v_i)$ $\leftarrow$ $\Delta b(v_i)$ + $[\sigma^{(v_i)}_{AP(v_m,v_t)} \cdot (x_t / x_t + 1)]$\label{algol:end_compute_reward} 
        \EndFor
        \State $p$ $\leftarrow$ $-\Delta b(v_i) \cdot \hat{d}_i$ \label{algol:start_reward_ratio_check}
        \If{$\hat{d}_i$ $\geq$ 1} 
            \State Insert $i$ into $Q$ with priority $p$ 
        \EndIf \label{algol:end_reward_ratio_check}
    \EndFor \label{algol:end_iteration}
\EndWhile
\end{algorithmic}
\end{algorithm}

Line \ref{algol:initial_cost} computes the number of deployable decoys $\hat{d}_m$ on each microservice defined as the available resources on the assigned node divided by the microservice resource consumption. This value is used to weight each microservice in the priority queue $Q$ according to the number of APs traversing them in lines  \ref{algol:initial_priority}-\ref{algol:first_queue_insertion}.
We allocate a decoy on the microservice with the highest priority in $Q$ (i.e. the one in the first position) by updating the AG topology in line \ref{algol:deploy_decoy} and we recompute the available resources for decoys in line \ref{algol:update_resources}. We iteratively calculate the number of generated DAPs in lines  \ref{algol:start_sideffect}-\ref{algol:end_sideffect}. In detail, line \ref{algol:update_betweenness_1} updates the BC values for other microservices in AG given the new decoy allocation, while line \ref{algol:update_gains_1} takes into consideration the total number of DAPs, denoted as $\theta^{(v_i)}_{v_t}$, generated by the interaction with the newly allocated decoy and the previously allocated ones. Lines \ref{algol:queue_clear}-\ref{algol:update_ratios} clean the queue and recompute the number of deployable decoys. These steps are needed to discard microservices that cannot be cloned as decoys, thus they are no longer considered as possible candidates. Lines \ref{algol:start_compute_reward}-\ref{algol:end_compute_reward} compute the priority of each microservice according to the potential number of DAPs, expressed as the BC increment $\Delta b(i)$, that can be generated by allocating an additional decoy on microservice $t$. These steps guide the decoy allocation computation in the next iteration as the algorithm is incentivized to select microservices with the highest BC gain. Lines \ref{algol:start_reward_ratio_check}-\ref{algol:end_reward_ratio_check} insert the microservices into the queue with the updated priority only if the resources available are sufficient. The algorithm convergence is completed when $Q$ is empty, which indicates that the resources reserved for the decoys are exhausted.

The overall algorithm time complexity can be computed as follows. We restrict the analysis for a single iteration, referred by lines \ref{algol:extract_from_queue}-\ref{algol:end_iteration}, in order to provide the time complexity required to allocated a single decoy into the infrastructure. The most computationally expensive operation consists in calculating the number of DAPs that are generated by the previously allocated decoys. In this scenario, assuming that the priority queue $Q$ is implemented using a binary heap, these steps have a complexity of $\mathcal{O}(M \cdot (M + M) + M \cdot (M + \text{log}M) )$, which can be further simplified to $\mathcal{O}(M^2)$. Since the total number of iterations depends on the number of allocated decoys $D$, the overall algorithm complexity is $\mathcal{O}(D \cdot M^2)$.


\section{Performance evaluation}
We evaluate the performance of the presented deception mechanism under different configurations of number of active microservices and decoy resource ratio.
We compare the performance of the \textit{Optimal} decoy allocation defined in \eqref{eq:obj_func}-\eqref{eq:const_x}, which provides the performance upper-bound of the considered metrics, and the related \textit{Heuristic} allocation presented in Algorithm \ref{alg:heuristic} with the following three schemes:
\begin{itemize}
    \item{\textit{Linear}: this scheme provides a decoy allocation that maximizes the number of DAPs without accounting for the additional paths introduced by the decoys and \hl{follows a similar approach like the one proposed in} \cite{jin2019dseom}. For example, with reference to Fig. \ref{fig:DAP_generation}, this scheme would ignore the DAP between $d_{m}^{(1)}$ and $d_{m'}^{(1)}$. In other words, this scheme neglects the AG topology modifications due to the decoy placement. The solution is computed by formulating a linear integer optimization problem having $\bar{\sigma}^{(m)}_{AP(v_s,v_t)} $ as objective function instead of \eqref{eq:obj_func}. Formally, we have 
    \begin{equation}
       \max_{\mathbf{x}} \sum_{m \in M} x_{m} \cdot \sum_{v_s \neq v_m \in V} \sum_{v_t \neq v_s \in V} \bar{\sigma}^{(v_m)}_{AP(v_s,v_t)}
       \label{eq:obj_func_2}
    \end{equation} 
    subject to \eqref{eq:const_cpu}-\eqref{eq:const_x}}. Note that this formulation is still NP-hard as it can be reduced to the classical Knapsack problem. We assess the performance of this approach in order to analyze the performance gain provided by the \textit{Optimal} solution which accounts for every DAPs introduced by the decoy allocation. 
    \item{\textit{Sidecar}: this scheme is a simple decoy placement strategy that allocates decoys on the most vulnerable assets. In our scenario, we identify the most vulnerable microservices as those having the lowest weight $w$ associated to the in-ward edges in the AG. To provide a fair comparison with other schemes, which are resource-aware, we account for the resource consumption of the decoys by selecting the microservices to clone according to the priority metric of the heuristic algorithm. \bc{This approach computes the decoy allocation based on a local assessment of the vulnerability scores associated with each microservice and neglects whether those vulnerabilities are actually exploited by the various APs. Consequently, by comparing the \textit{Optimal} and \textit{Heuristic} schemes with this approach, we can better highlight any performance gain obtained by instead using the betweenness centrality to globally evaluate the vulnerability of each microservice in terms of the number of APs traversing it.}}
    
    \item{\textit{Random}: this scheme randomly allocates the decoys and it is used as a performance lower bound.}
\end{itemize}

 \hl{We define the following metrics to compare the security performance achieved by the aforementioned decoy allocation schemes}:
\begin{itemize}
    \item{\textbf{Decoy interaction probability}}: \hl{this is the probability that an attacker interacts with a decoy when moving laterally between the microservices composing an AP. In other words, this is the probability that an attacker follows a DAP in order to reach its target. Formally, we compute the decoy interaction probability associated to a decoy placement configuration} $\mathbf{x}$ as
    \begin{equation}
        P_{DAP}(\mathbf{x}) = \frac{N_{DAP}}{N_{AP}+N_{DAP}},
        \label{eq:decoy_interaction_prob}
    \end{equation}
    \hl{where $N_{AP}$ is the initial number of APs in the attack graph without any decoy, and $N_{DAP}$ is the number of DAPs generated on the attack graph by the decoy placement configuration $\mathbf{x}$. We define this metric in order to provide a more practical insight about the deception performance of the considered schemes as it measures the decoys likelihood to intercept attacks.}
    \item{\textbf{Average number of decoy per AP}}: \hl{this metric measures the average ratio of decoys over the number of legitimate microservices contained in each AP. We define this indicator to evaluate the efficiency of the various schemes in condensing the allocation of decoys on microservices that are traversed by an AP. Intuitively, the higher is this value, the higher is the frequency that an attacker can interact with more than one decoy before reaching its target. As a result, the defender can collect more information about the attacker techniques after each decoy interaction. Formally, we compute the average decoy number per AP associated to a decoy placement configuration} $\mathbf{x}$ as
    \begin{equation}
        D_{AP}(\mathbf{x}) = \frac{1}{N_{AP}}  \sum_{v_s \in V} \sum_{v_t \neq v_s \in V}\frac{n_{AP(v_s,v_t)}}{|AP(v_s,v_t)|},
        \label{eq:avg_decoy_number_per_AP}
    \end{equation}
    \hl{where $n_{AP(v_s,v_t)}$ is the number of decoy allocated on the AP defined by the vertex pair $(v_s,v_t)$.}
\end{itemize}

\subsection{Simulation setup}

The simulation environment was developed in Python and \bc{executed on Apple M1 Pro with 14-cores 3.2 GHz CPU and 16GB RAM}. In particular, the decoy deployment as well as the AG related computations were implemented using \textit{NetworkX} library \cite{networkx}, which provides many functionalities to handle operations in graphs. The optimal and linear approaches were computed using the \textit{SCIP} optimization suite \cite{SCIPv8}.

Due to the impossibility to retrieve practical example of microservice-based applications, we synthetically generated the microservice diagram topology according to a \textit{Barabasi-Albert} graph model as already proposed in \cite{podolskiy2020weakest}. These graphs, composed by few highly-connected vertices and many low-connected vertices \cite{yook2002modeling}, resemble the typical structure of the communication flows within microservice-based applications where few microservices dispatch tasks toward multiple microservices that are rarely connected with one another. 


To effectively model a realistic resource profile for a given microservice deployment, we employed a public dataset containing the traces of the normalized computing nodes capacities and the resource consumption of the allocated containers in the Alibaba cloud infrastructure \cite{alibaba_traces_2017}. We generated the resource deployment configuration matrices $\mathbf{R}^{(cpu)}$ and $\mathbf{R}^{(ram)}$ by randomly sampling a number of containers corresponding to the number of considered microservices. In other words, we mapped the resource consumption of Alibaba containers and the related allocation on the computing nodes as $r_{m,n}^{(ram)}$ and $r_{m,n}^{(cpu)}$. We chose the number of computational nodes $N$ by progressively deploying each active microservice on the same node up to the $70\%$ of its capacity before selecting a new one. For example, by following this procedure, the number of nodes required to accommodate 500 microservices, which is the maximum amount considered in the simulations, is $N = 20$. Based on the resulting microservice deployment, we reserved a share of the remaining resources in each computing node for the allocation of decoys according to the decoy resource ratio $\delta$. We generated the edge weights in the AG by first assigning a random number of vulnerabilities between 3 to 5 to each microservice, i.e. $|\mathcal{V}_m| \in \{3,4,5\}, \forall m \in M$. Then, for each vulnerability of every microservice, we extracted random scores of $EM$ and $ECM$ using the  CVSS calculator tool \cite{cvss2019common} in order to compute the final value of $w$  with \eqref{eq:weight}. 

\bc{Considering the largest microservice configuration used in the simulations, i.e., $M=500$, the AG topologies generated according to the aforementioned procedure had the following graph metrics values: the average vertex degree was 25, the average shortest path length was 4 and the overall graph density, which is the ratio between the actual number of edges and the largest number of possible edges, was 0.05. Intuitively, these values indicate graph topologies formed by multiple clustered components of highly connected vertices. In particular, the vertices belonging to the same cluster correspond to microservices allocated on the same computing node since they can be exploited by an attacker to compromise any other co-located microservice using container escape strategies. This dynamic inflates the vertex degree while lowering the attack path length (which in our case is also the shortest path) as any microservice belonging to the same computing node can be compromised with a one-step lateral movement. Differently, in order to laterally move between microservices on different nodes, the attacker can leverage, for instance, the available communications interfaces defined by the application. This effect causes the sparsity of the resulting graph as microservices belonging to different nodes preserve their characteristic of being rarely connected.} 

Finally, we randomly sampled 100 configurations of microservice architectures and resource deployment and we plotted the average results within the 95\% confidence intervals for each considered simulation instance. \bc{ We resume the main simulation parameters in Table \ref{tab:simulation_parameters}.}

\begin{table}[t]
	\centering
	\scriptsize
	\caption{\bc{Simulation parameters}}
	\renewcommand{\arraystretch}{1.2}
	\begin{tabular}{|p{3.7cm}|c|}
        \hline
        Maximum CPU per computing node & 96 cores \\ 
        \hline
        Maximum RAM per computing node & Normalized to 100 \\
        \hline
        Number of microservices & $M \in [100,500]$ \\
        \hline
        Microservice CPU request  & $\bar{r}_{m}^{(cpu)}  \in [1.0,3.0] $ cores \\
        \hline
        Microservice RAM request & $\bar{r}_{m}^{(ram)}  \in [0.52,2.3] \% $ \\
        \hline
	\end{tabular}
    \label{tab:simulation_parameters}
\end{table}

\begin{figure}[t]
    \centering
    \includegraphics[width=1\linewidth]{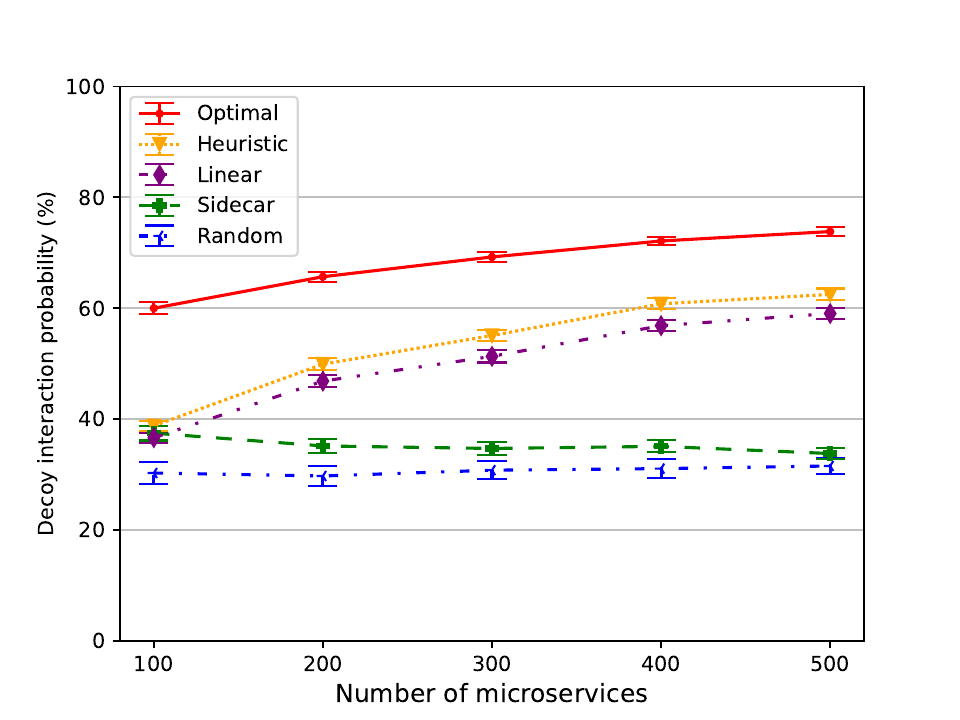}
    \caption{Decoy interaction probability $P_{DAP}$ when the number of microservices is increased from $M = 100$ to $M = 500$. The decoy resource ratio is $\delta = 0.3$.}
    \label{fig:microservices_paths_covered}
\end{figure}

\begin{figure}[t]
    \centering
    \includegraphics[trim={0 0 0 1cm},clip=true,width=1\linewidth]{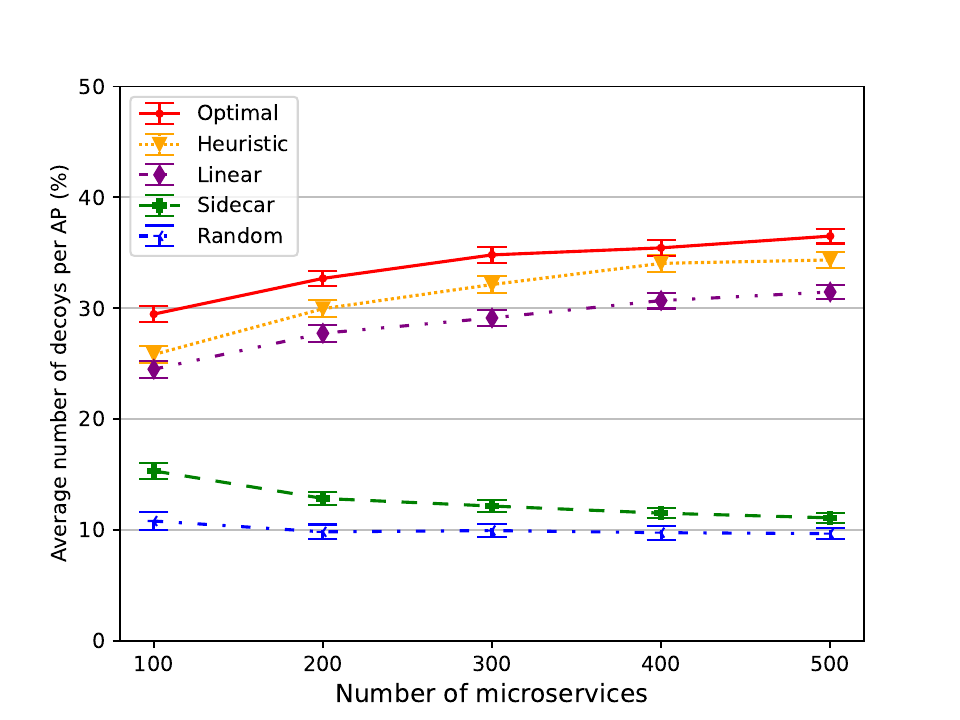}
    \caption{Average number of decoy per AP ${D}_{AP}$ when the number of microservices is increased from $M = 100$ to $M = 500$. The decoy resource ratio is $\delta = 0.3$.}
    \label{fig:microservices_avg_decoys_per_ap}
\end{figure}

\begin{figure}[t]
    \centering
    \includegraphics[width=1\linewidth]{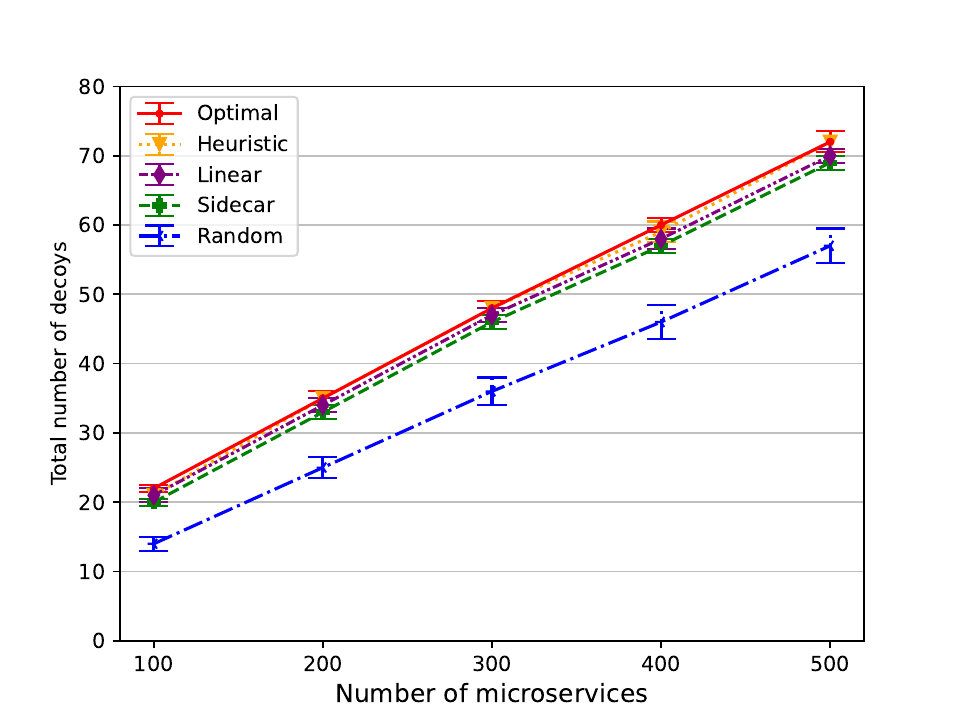}
    \caption{Total number of allocated decoys when the number of microservices is increased from $M = 100$ to $M = 500$. The decoy resource ratio is $\delta = 0.3$.}
    \label{fig:microservices_perc_decoys}
\end{figure}

\begin{table}[th]
	\centering
	\scriptsize
	\caption{\bc{Decoy-to-microservice ratio ($\%$)}}
	\vspace{-2mm}
	\label{tab:decoy_ratio}
	\renewcommand{\arraystretch}{1.2}
	\begin{tabular}{|l|c|c|c|c|c|}
        \hline
         M & Optimal & Heuristic & Linear & Sidecar & Random \\
		\hline
		$100$ & $21.50\%$ & $21.12\%$ & $20.83\%$ & $20.35\%$ & $15.39\%$ \\
		\hline
		$200$ & $17.70\%$ & $17.38\%$ & $17.06\%$ & $16.58\%$ & $12.27\%$ \\
		\hline
		$300$ & $16.08\%$ & $15.94\%$ & $15.56\%$ & $15.21\%$ & $12.08\%$ \\
        \hline
        $400$ & $15.07\%$ & $14.85\%$ & $14.60\%$ & $14.26\%$ & $11.58\%$ \\
        \hline
        $500$ & $14.30\%$ & $14.34\%$ & $13.97\%$ & $13.74\%$ & $11.47\%$ \\
		\hline
	\end{tabular}
 \end{table}

\begin{figure}[t]
    \centering
    \includegraphics[width=1\linewidth]{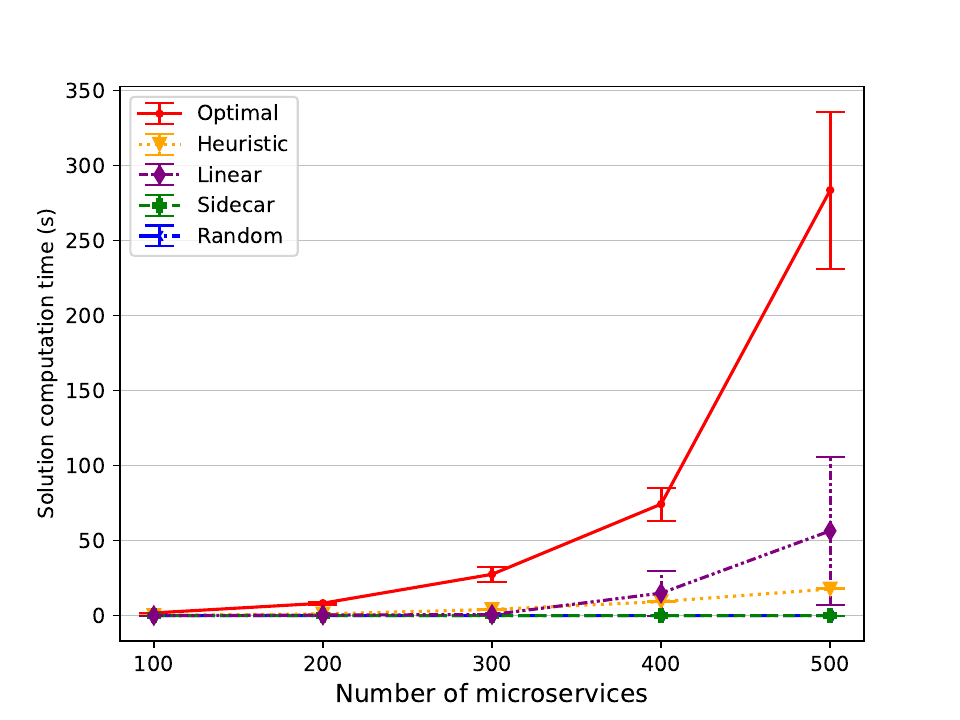}
    \caption{Computational complexity of the considered schemes when the number of microservices is increased from $M = 100$ to $M = 500$. The decoy resource ratio is $\delta = 0.3$.}
    \label{fig:microservices_time}
\end{figure}

\subsection{Variable number of microservices}
We assess the results by fixing the decoy resource ratio as $\delta = 0.3$ while increasing the number of microservice from $M=100$ to $M=500$.


In Fig. \ref{fig:microservices_paths_covered} we present the decoy interaction probability as defined in \eqref{eq:decoy_interaction_prob}.
The optimal solution achieves the best performance since it spreads more effectively the decoys within the current microservice deployment. In detail, the optimal formulation selects microservices that are traversed by a high number of APs and, at the same time, by a high number of DAPs which are possibly generated by already deployed decoys given the resource available. 
The heuristic scheme provides some notable optimality gap since its greedy approach prioritizes the allocation of decoys along the same AP instead of diversifying the allocation on other APs. However, it still generates a higher number of DAPs compared to the linear scheme.
The latter underestimates the number of generated DAPs as it neglects the impact of the allocated decoys on the AG topology.  This decoy placement configuration does not take advantage of DAPs generated between decoys to increase the possible number of intercepted attacks, hence it suffers from an evident performance loss compared to the optimal approach.
The sidecar is outperformed by all schemes since it does not consider the sequence of microservices that must be violated by the attacker in order to reach the target. In detail, highly vulnerable microservices are reached by few APs when they are surrounded by more secure microservices. Consequently, the allocated decoys generates a lower number of DAPs. Moreover, by increasing the number of microservices, this effect is exacerbated as there are more microservices acting as “filters" that further reduce the number of APs traversing the selected microservices. 


In Fig. \ref{fig:microservices_avg_decoys_per_ap} we show the average number of decoys per AP as defined in \eqref{eq:avg_decoy_number_per_AP}. 
In general, the discussion of the previous plot also applies for this scenario. The optimal scheme ensures the highest number of decoys per AP as it places the decoys along microservices contained in multiple APs. The heuristic scheme achieves similar performance and provides better gain compared to the linear and sidecar schemes. In particular, the greedy nature of the heuristic scheme encourages the allocation of decoys on microservices within the same AP. This strategy generates a high amount of DAPs at each new algorithm iteration thanks to the already allocated decoys along that AP in the previous steps. 


In Fig. \ref{fig:microservices_perc_decoys} we show the total number of allocated decoys. \bc{Moreover, in Table \ref{tab:decoy_ratio}, we also report the ratio of decoys over the current number of deployed microservices, referred to as decoy-to-microservice ratio.} The number of decoys increases as more computing nodes are activated to accommodate the required microservices. 
Generally, excluding the random scheme which is resource-agnostic, the various schemes roughly deploy the same number of decoys for each microservice deployment configuration. 
This behavior suggests that the performance gain provided by the optimal and heuristic schemes derive from a decoy allocation that is tailored to the existing APs traversing the various microservices and it is not due to a higher number of allocated decoys that artificially inflates the number of generated DAPs. \bc{As shown in Table \ref{tab:decoy_ratio} the decoy-to-microservice ratio is approximately $21\%$ for $N=100$ and $14\%$ for $N=500$ for the optimal and heuristic schemes. Therefore, this trend indicates that the number of decoys increases more slowly than the number of active microservices. Nonetheless, in contrast to the sidecar scheme, such modest decoy increment still allows the optimal and heuristic decoy allocations to increase their APs interception performance.}


\begin{figure}[t]
    \centering
    \includegraphics[width=1\linewidth]{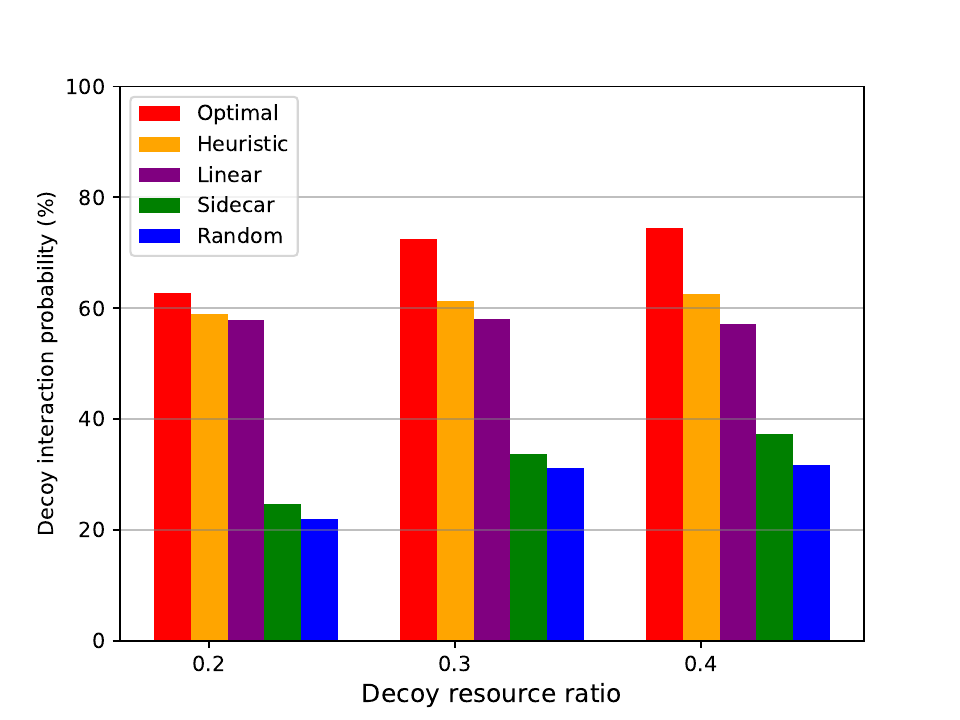}
    \caption{Decoy interaction probability $P_{DAP}$ when the decoy resource ratio is  $\delta = \{0.2, 0.3, 0.4\}$. The number of active microservices is $M = 500$.}
    \label{fig:resources_paths_covered}
\end{figure}

\begin{figure}[t]
    \centering
    \includegraphics[width=1\linewidth]{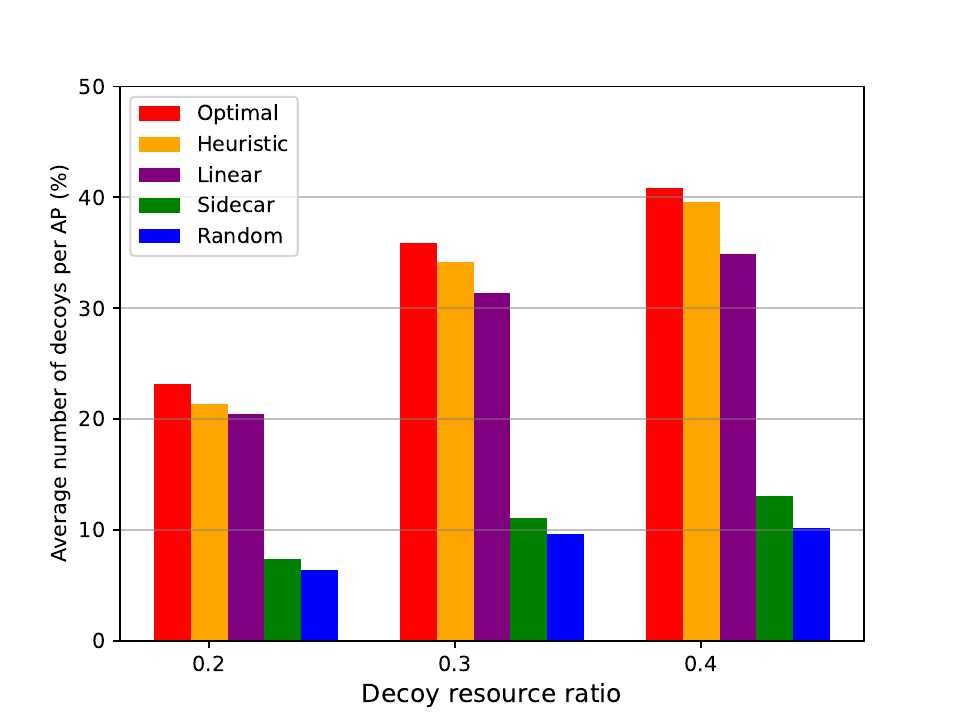}
    \caption{Average number of decoy per AP $D_{AP}$ when the decoy resource ratio is  $\delta = \{0.2, 0.3, 0.4\}$. The number of active microservices is $M = 500$.}
    \label{fig:resources_avg_decoys_per_aps}
\end{figure}

In Fig. \ref{fig:microservices_time} we show the convergence performance of the considered schemes. \bc{Note that the number of computing nodes scales according to the number of active microservices in order to ensure a maximum resource utilization in each node of $70\%$, as mentioned in the simulation setup subsection. Therefore, the computed resolution time for each microservice configuration also includes the overhead introduced by the increasing amount of computing nodes}. The non-linearity of \eqref{eq:obj_func} makes the optimal scheme unsuitable for large-scale microservice deployments that need to be frequently updated due to the non-negligible convergence time. \bc{In other words, any modification of the deployment requirements associated with the legitimate microservices causes a reconfiguration of the current decoy allocation.  For example, the resizing of microservice resources to accommodate different workload conditions, the deployment of additional microservices to provide new applications or the migration of microservices to reduce resource contention issues in overloaded computing nodes are events triggering a decoy placement recomputation.} Conversely, the heuristic decoy allocation requires a considerably lower computational complexity, thus it is a preferable option to handle the aforementioned scenarios. Moreover, the proposed heuristic also outperforms the linear approach, that suffers from a similar trend as the optimal scheme when the number of microservices is higher than 300. The sidecar scheme is characterized by a very low complexity due to the simplicity of its approach, which however achieves poor results in terms of DAP generation as previously discussed. As a consequence, the optimal and heuristic schemes provide the most effective solutions to ensure an effective deception strategy and a scalable decoy placement configuration, respectively.


\subsection{\bc{Variable decoy resource ratio}}

We present the performance of the considered decoy allocation schemes for different configurations of decoy resource ratio $\delta$ in the scenario of $M = 500$.


In Fig. \ref{fig:resources_paths_covered}, we show the decoy interaction probability for different resource configurations. By increasing the number of resources assigned to the decoy placement, the optimal and heuristic schemes increase the APs coverage by allocating a higher number of decoys. Conversely, we observe the performance degradation of the linear scheme due to the increasing  underestimation error of the DAPs number that gets more severe as the decoy deployment availability increases. 

In Fig. \ref{fig:resources_avg_decoys_per_aps}, we show the average number of decoys per AP. The heuristic scheme performance is comparable to the optimal scheme in every configuration. This result is achieved by distributing the decoy allocation on the available nodes in order to balance the resource consumption. In detail, by ranking microservices according to the number of deployable decoys, the heuristic scheme prioritizes the placement of decoys that can produce a high number of DAPs and, at the same time, have a low resource consumption.



\begin{figure}[t]
    \centering
    \includegraphics[width=1\linewidth]{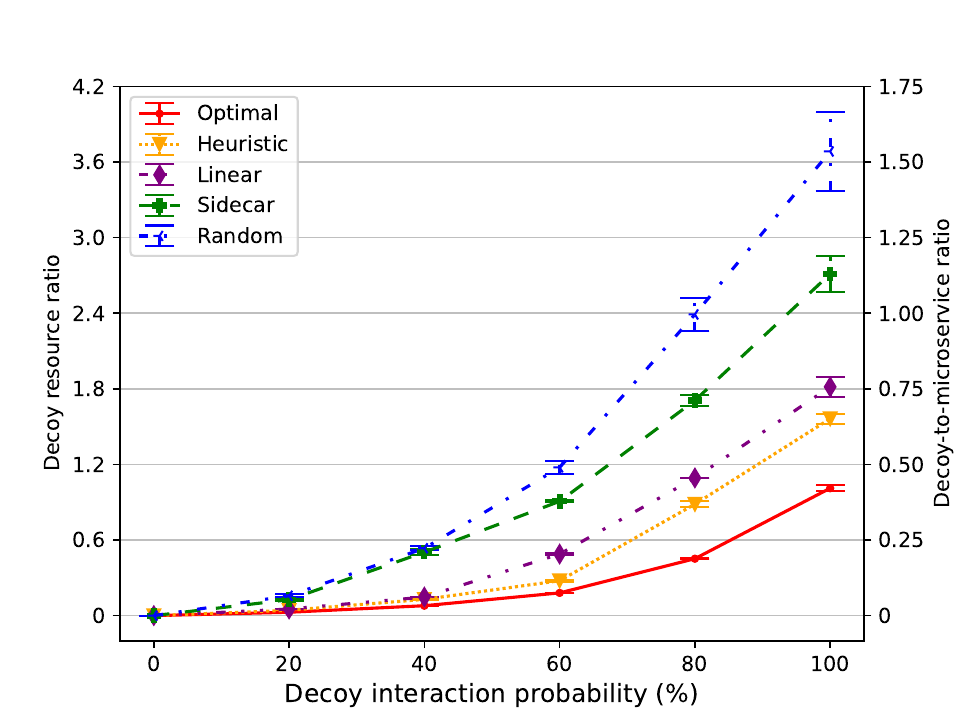}
    \caption{\bc{Average decoy resource ratio $\delta$ (left side y-axis) with corresponding decoy-to-microservice ratio (right side y-axis) for different values of decoy interaction probability. The number of active microservices is $M = 500$.}}
    \label{fig:ap_coverage_vs_delta}
\end{figure}

\bc{In Fig. \ref{fig:ap_coverage_vs_delta}, we provide additional insights about the resource efficiency performance of the various schemes by evaluating the required decoy resource ratio (i.e., $\delta$) and the corresponding decoy-to-microservice ratio to obtain a specific decoy interaction probability. In other words, the plot shows what is the fraction of available computational resources that must be reserved for each scheme in order to ensure that the corresponding decoy allocation can intercept an AP with a given probability. 
For example, assuming a target decoy interaction probability of $80\%$, the optimal scheme can satisfy this requirement using $50\%$ of the available resources (i.e., $\delta = 0.5$) to clone $20\%$ of the currently deployed microservices as decoys. In contrast, for the same decoy interaction probability value, the sidecar scheme would roughly require $160\%$ of the available resources. In a practical scenario, such decoy resource ratio value implies the necessity to increase the number of computing nodes in order to accommodate all the decoys as the current available node capacity is not sufficient. Regardless of the specific value of $\delta$, which is inherently related to the resource availability in each computing node, the overall trend of the optimal and heuristic schemes showcases their resource efficiency performance compared to the other approaches. In particular, they consistently require the least amount of resources to achieve the same decoy interaction probability. By accurately modeling the DAPs generation dynamic, the proposed solutions can effectively exploit the available resources to maximize the decoys interception performance.}

\subsection{Sensitivity to attacker's behavior}

\begin{figure}[t]
    \centering
    \includegraphics[width=1\linewidth]{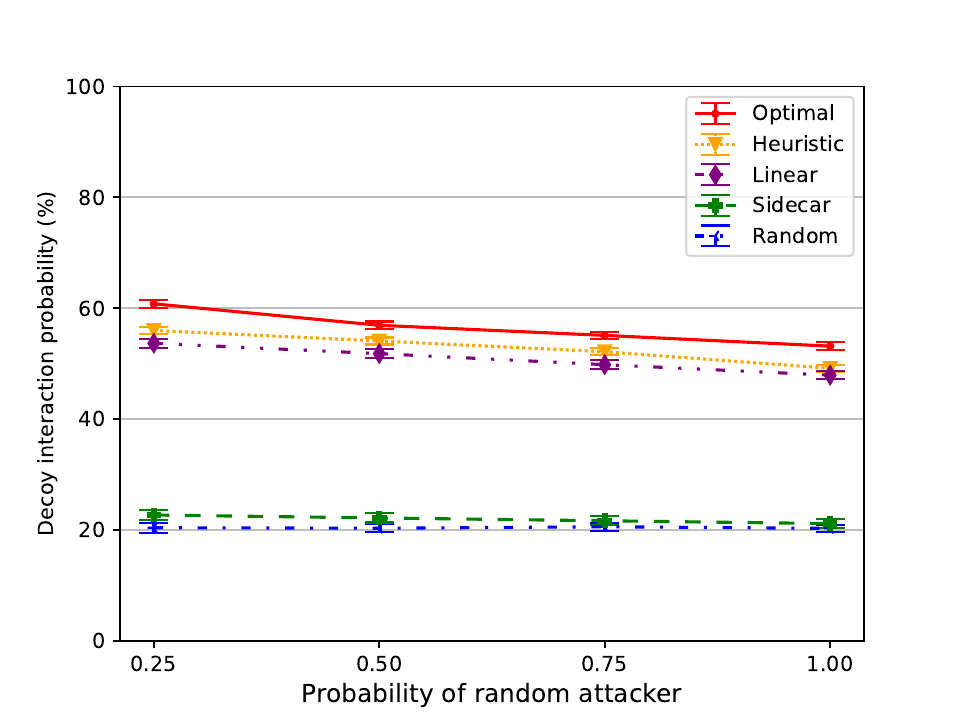}
    \caption{Decoy interaction probability $P_{DAP}$ as the probability of a random attacker increases. The decoy resource ratio is  $\delta = 0.3$. The number of active microservices is $M = 500$.}
    \label{fig:sensitivity_decoy_inter_prob}
\end{figure}

As final result, we investigate how the decoy interaction probability metric responds to various attacker profiles. We simulate a random attacker behavior that differs from the ``least impedance" AP assumption outlined in Section 3. To achieve this, for every source and target pair in AG, we randomly sample an AP with probability of $p$. Conversely, we retain the AP defined as the shortest path in AG with a probability of 1-$p$. Subsequently, we evaluate the decoy interaction probability using the same deployment of decoys that maximizes the number of DAPs for an attacker that strictly follows the shortest path in the AG.

In Fig. \ref{fig:sensitivity_decoy_inter_prob}, we show the outcomes of our analysis by plotting the decoy interaction probability as the probability of introducing random AP (i.e., $p$) gradually increases. Specifically, when such probability is equal to 1, the attacker behavior is fully random, hence the selected AP is unlikely to match the shortest path. With respect to the results, besides the random scheme, which is unaffected by the attacker profile, we observe a performance degradation of the optimal and heuristic solutions that still, however, outperform the linear scheme. This behavior is expected as the decoy placement maximizes the number of intercepted APs, which now differs when alternative APs are considered. Nonetheless, the performance loss is modest and can be neglected compared to the considerable gains achieved over the sidecar scheme. 
The resilience of the proposed deception strategy to various attacker profiles can be attributed to the all-pair shortest path approach used for configuring decoy placements. In particular, our approach considers all source and target pairs within AG. Consequently, some microservices belonging to the predicted AP, when cloned as decoys, may also intercept random APs, as they share sub-path segments. Thanks to this dynamic, the resulting decoy placement configuration remains effective even when dealing with attacker behaviors that deviate from the worst-case scenario, providing a reliable cyber-deception solution.

\section{Conclusion}
Cyber deception is a promising defense strategy to enhance the security of microservice architectures. By deploying deceptive microservices within the infrastructure, it is possible to intercept attacks and learn about attackers' TTPs with a limited complexity overhead.   
In light of this, we considered the problem of a resource-aware decoy allocation strategy under a limited budget of available computational resources. We modeled the attacker lateral movements between microservices using graph theory, where we considered an attack path as the sequence of microservices violated by the attacker in order to reach its target. We proposed an analytical formulation to quantify the number of deceptive attack paths generated by an arbitrary decoy placement, and used this expression to design an integer non-linear optimization problem that maximizes the number of deceptive attack paths given the available resources. We also designed a low-complexity heuristic decoy allocation scheme to approximate the optimal solution. In the performance evaluation, our approach outperformed benchmark schemes on several metrics while using the same number of decoys. 

\bc{As future work, we plan to augment the proposed cyber deception approach by combining it with moving target defense strategies in order to alter the attacker’s perception of the defender environment. For example, the flexibility offered by cloud-native environments in deploying containerized microservices can be exploited to dynamically reconfigure the current decoy allocation according to the attacker's actions. This adaptive strategy can be used to waste the attacker's resources and slow his progress toward the objective.}

\section*{Acknowledgements}
This work has received funding from the EU Horizon Europe R\&I Programme under Grant Agreement no. 101070473 (FLUIDOS).

\bibliographystyle{IEEEtran}
\bibliography{bibliography}

\end{document}